\documentclass{ptephy_v1}%%%%%% to generate preprint number with ptep logo

\usepackage{amsmath}% for dealing with mathematics,
\usepackage{tikz}

\begin{document}

\title{Study of the neutrino-oxygen cross sections of the charged-current reaction $^{16}$O($\bar{\nu}_e, e^+$)$^{16}$N(0 MeV, $2^-$) and the neutral-current reaction $^{16}$O($\nu, \nu^{\prime}$)$^{16}$O(12.97/12.53 MeV, $2^-$), producing  high-energy $\gamma$  rays }

%%%% To generate auto affiliation numbers please use \author{}\affil{} command

\author[1,*]{Makoto \textsc{SAKUDA}}
\affil[1]{Physics Department, Okayama University, Okayama 700-8530, Japan}

\author[2,3*]{Toshio \textsc{SUZUKI}}
\affil[2]{Department of Physics, College of Humanities and Sciences, Nihon University, Tokyo 156-8550, Japan}
\affil[3]{NAT Research Center, NAT Corporation, 3129-45 Hibara Muramatsu, Tokai-mura, Naka-gun, Ibaraki 319-1112, Japan}
%\author[3]{Mandeep Singh \textsc{REEN}}
%\affil[3]{Department of Physics, Akal University, Punjab 151302, India}

\author[4,*]{Ken'ichiro \textsc{NAKAZATO}}
\affil[4]{Faculty of Arts and Science, Kyushu University, Fukuoka 819-0395, Japan}

\author[5]{Hideyuki \textsc{SUZUKI}}
\affil[5]{Department of Physics, Faculty of Science and Technology, Tokyo University of Science, Chiba 278-8510, Japan}

\affil[*]{\email{sakuda-m@okayama-u.ac.jp, suzuki.toshio@nihon-u.ac.jp}, nakazato@artsci.kyushu-u.ac.jp}
%\author[3]{Insert fourth author name here} %%% Use optional bracket [3] to change the respective address
%\affil{Insert third author address here}

%\author{Insert last author name here\thanks{These authors contributed equally to this work}}
%\affil{Insert last author address here}

%%% To include the collaborator name... Please use the command "\collaborator"
%%% For example: \collaborator{ATLAS Collaboration}

\begin{abstract}%
In the previous work, we discussed the cross section and the detection of 4.4-MeV $\gamma$ rays produced in the neutrino neutral-current (NC) reaction $^{16}$O($\nu, \nu^{\prime}$)$^{16}$O(12.97 MeV and 12.53 MeV, $2^-$) in a water Cherenkov detector at the low energy below 100 MeV. In this report, we further investigated both the charged-current (CC) reaction $^{16}$O($\bar\nu_e, e^+$)$^{16}$N(0 MeV, $2^-$) and  the NC reaction$^{16}$O($\nu, \nu^{\prime}$)$^{16}$O(12.97 MeV and 12.53 MeV, $2^-$), producing high-energy $\gamma$  rays, in which the more solid identification of the reactions can be applied via the coincidence method. 
%We also evaluate the number of such events induced by neutrinos from supernova explosion which can be observed by the Super-Kamiokande, a 32 kton water Cherenkov detector in the Earth.
\end{abstract}

\subjectindex{C43, D02, D03, D21, F22}

\maketitle

\section{Introduction}

The 12.97-MeV and 12.53-MeV states are the first strong $2^-$ excited states of $^{16}$O just above the proton separation energy (12.1 MeV). 
The 12.97-MeV state, which is nearly an isospin $T$=1 state, is one of the dominant multipoles in the neutrino-oxygen interactions at low energy below 100 MeV. 
The electromagnetic form factors $F^2(q)$ of these states were measured in ($e,e^{\prime}$) reactions in 1960~\cite{Vanpraet,Stroetzel, Kim, Sick}. 
%%The reduced transition probabilities $B(M2,q)$ of those states were also measured by Stroetzel  and Kim ${\it et\ al.}$~\cite{Stroetzel, Kim}.  
No new measurements of those states in ($e,e^{\prime}$) reactions have been performed since then. 
Donnelly and Walecka~\cite{Donnelly1,Donnelly2,Walecka75} calculated the neutrino-$^{16}$O cross sections at $E_x$=12-20 MeV  precisely with accuracy of 15-20\% after they analysed the data of $^{16}$O$(e, e^{\prime})^{16}$O$(E_x$=12-20 MeV) scattering and semi-leptonic weak interactions (muon capture and $\beta$ decay) and evaluated the reduction factors ($a/\xi$=0.6-0.7) to the transition amplitudes of their model. This reduction in transition amplitudes of a calculation model (or in the coupling constant) is sometimes called a quenching factor. 
Haxton~\cite{Haxton} calculated the cross sections of the charged-current (CC) neutrino-oxygen $^{16}$O($\nu_e$, e$^-$)$^{16}$F and $^{16}$O($\bar{\nu}_e$, e$^{+}$)$^{16}$N reactions, using the quenching factors for negative parity states, which  were evaluated in Ref.~\cite{Donnelly1,Donnelly2,Walecka75}. He further examined the CC cross sections to the bound states (2$^-$, 0$^-$, 3$^-$, and 1$^-$) of $^{16}$N, which are followed by the $\beta^-$ decay to the ground state or  the excited state (6.13 MeV) of $^{16}$O. The total energy given by $\beta^-$ and $\gamma$ ray (6.13 MeV) was estimated to be about 8 MeV.  He concluded that since the CC cross section to the bound states of $^{16}$N never exceeds  1\% of the dominant inverse beta decay (IBD) reaction in supernova neutrino bursts,  the extra delayed signal of 8 MeV scattered over the decay time ($T_{1/2}$=7.13 sec)  gives only negligible effect on the event timing,  which is determined by the dominant IBD events. 

At the time of these analysis, the isospin mixing of the two 2$^-$ states at 12.53 MeV and 12.97 MeV was not known and was not considered. 
There have been several reports on the isospin mixing between the 12.97 MeV ($T=1$) and 12.53 MeV ($T=0$) states previously~\cite{Stroetzel, Wagner, Leavitt, Zijderhand, Charity}. 
These physical two $2^-$ states (the higher energy state $|U\rangle$ and the lower energy state $|D\rangle$) are written in terms of the pure isospin states as,  
\begin{eqnarray}\label{isospin}
|U \rangle &=& \sqrt{1-\beta ^2} \,|U, T=1 \rangle -\beta \,|U,T=0 \rangle, \nonumber  \\ 
|D \rangle &=& \sqrt{1-\beta ^2} \,|D, T=0 \rangle +\beta \,|D,T=1 \rangle,
\label{eq:ud-states}
\end{eqnarray}  
where $\beta$ is the isospin mixing parameter. A well-known example of the isospin mixing is that between the two excited states of $^{12}$C at 12.71 MeV ($1^+, T=0$) and 15.11 MeV ($1^+, T=1$)~\cite{Adelberger, Flanz, Cosel}. 

In our previous work~\cite{MSakuda}, we followed the analysis by Donnelly and Walecka, evaluating both the quenching factor $f_s=g_s^{{\rm eff}}/g_s$ of the  spin $g$ factor and the isospin-mixing parameter $\beta$ of the two $2^-$ states  to be $f_s=0.65\pm 0.05$ and $\beta=0.25\pm 0.05$, respectively, and also  determining the quenching factor $f_A=g_A^{{\rm eff}}/g_A$ of the axial-vector coupling constant to be $f_A=0.68\pm 0.05$. Then, we discussed the cross section of 4.4-MeV $\gamma$ ray production in the neutrino neutral-current (NC) reaction $^{16}$O($\nu, \nu^{\prime}$)$^{16}$O(12.97 MeV, $2^-$) in a water Cherenkov detector at the low energy below 100 MeV.  

The Super-Kamiokande (SK) experiment summarizes the following three detection channels from supernova (SN) neutrino bursts as described in Ref.~\cite{SK-SN, SK-SNwatch}: (1) the IBD reaction $p(\bar{\nu}_e,e^+)n$, (2) the neutrino-electron elastic scattering, and (3) the CC reactions, $^{16}$O($\nu_e, e^-$)$^{16}$F and $^{16}$O($\bar\nu_e, e^+$)$^{16}$N. The first IBD reaction is the main interaction channel, responsible for about 90\% of the reactions in water. The second one is a subdominant channel, useful for determining the direction of the SN. The third CC neutrino-oxygen interactions are also  subdominant ones. Their cross sections were calculated initially by the authors of Ref.~\cite{Haxton, Kolbe03} and recently in Ref.~\cite{TSuzuki-16O}, and the electron spectra of the  CC reactions were discussed by the authors of Ref.~\cite{NakazatoCC, SK-SNwatch, HyperK-SN}. 

The JUNO experiment~\cite{JUNO}, a Liquid Scintillator detector of 20-kton fiducial mass, uses the 15.11-MeV $\gamma$-ray emission of the NC reaction $^{12}$C($\nu, \nu^{\prime}$)$^{12}$C(15.11 MeV, $1^+$), and  the CC reactions,  $^{12}$C($\nu_e, e^-$)$^{12}$N(g.s., $1^+$) and $^{12}$C($\bar{\nu}_e, e^+$)$^{12}$B(g.s., $1^+$)~\cite{Fukugita}, as the main detection channels for the analysis of the SN neutrino bursts, in addition to the IBD reaction, elastic $\nu$–p scattering and elastic  $\nu$–e scattering~\cite{Strumia, Vissani1, Beacom}. We denote the ground state  as g.s. hereafter.

This study on the CC/NC neutrino-oxygen reactions using $2^-$ ($T=1$) states of $^{16}$N and $^{16}$O is motivated by the well-studied CC/NC neutrino-carbon reactions using $1^+$ ($T$=1) ground states of $^{12}$B, $^{12}$C and $^{12}$N, where  both CC and NC reactions of neutrino-$^{12}$C are already measured by the low-energy  neutrino experiments~\cite{KarmenNC1, KarmenNC2, KarmenCC1, KarmenCC2, LSNDCCe, LSNDnuee, E225, LSNDCCm1,LSNDCCm2}. We investigate both the CC $^{16}$O($\bar{\nu}_e$, e$^{+}$)$^{16}$N(g.s., 2$^-$) and its $\beta$ decay to $^{16}$O. Furthermore, we investigate the NC reaction cross sections from the two 2$^-$ states (12.97 MeV and 12.53 MeV) of $^{16}$O, producing high-energy $\gamma$ rays. We discuss a possible coincidence method to identify these CC and NC reactions unambiguously, which can be used not only in the SK experiment but also in the future Hyper-K experiment~\cite{HyperK-SN}. 

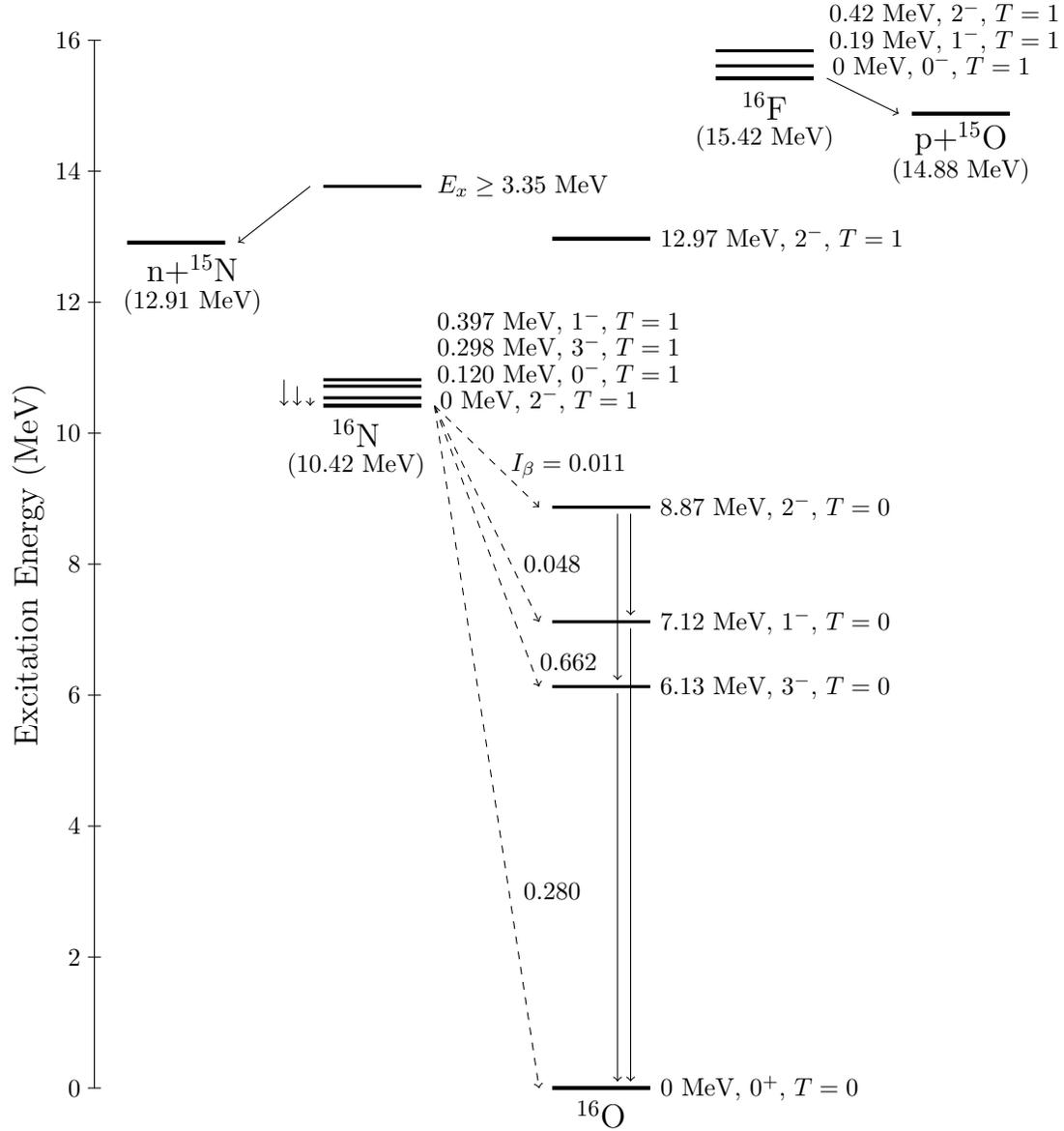
\begin{figure}[ht]
\begin{tikzpicture}[scale=0.9, every node/.style={scale=0.9}]
% Draw the energy axis
\draw (1,0) -- (1,16);
\node[rotate=90, font=\Large] at (0,8) {Excitation Energy (MeV)};
% Adding energy labels
\foreach \y in {0, 2, 4, 6, 8, 10, 12, 14, 16}
    \draw (0.9,\y) -- (1.1,\y) node[left=5pt] {\y};
% Draw the excitation levels for n+15N
\draw[->] (4.3,13.77) -- (3.2,12.91);
\draw[ultra thick] (1.5,12.91) -- (3,12.91);
\node[font=\Large] at (2.5,12.5) {n+$^{15}$N};
\node at (2.5,12) {(12.91 MeV)};
% Draw the excitation levels for 16N
\draw[very thick] (4.5,13.77) -- (6,13.77);
\node at (7.55,13.77) {$E_x \geq 3.35$ MeV};
\draw[very thick] (4.5,10.817) -- (6,10.817);
\draw[very thick] (4.5,10.718) -- (6,10.718);
\draw[very thick] (4.5,10.54) -- (6,10.54);
\draw[ultra thick] (4.5,10.42) -- (6,10.42);
\node at (8.1,11.7) {0.397 MeV, 1$^-$, $T=1$};
\node at (8.1,11.3) {0.298 MeV, 3$^-$, $T=1$};
\node at (8.1,10.9) {0.120 MeV, 0$^-$, $T=1$};
\node at (7.8,10.5) {0 MeV, 2$^-$, $T=1$};
\draw[->] (3.9,10.817) -- (3.9,10.42);
\draw[->] (4.1,10.718) -- (4.1,10.42);
\draw[->] (4.3,10.54) -- (4.3,10.42);
\node[font=\Large] at (5,10) {$^{16}$N};
\node at (5,9.5) {(10.42 MeV)};
\draw[dashed,->] (6.2,10.42) -- (7.8,8.87);
\draw[dashed,->] (6.2,10.42) -- (7.8,7.12);
\draw[dashed,->] (6.2,10.42) -- (7.8,6.13);
\draw[dashed,->] (6.2,10.42) -- (7.8,0);
\node at (8.25,9.5) {$I_\beta=0.011$};
\node at (8,8) {0.048};
\node at (8.25,6.5) {0.662};
\node at (8,3) {0.280};
% Draw the excitation levels for 16O
\draw[ultra thick] (8,0) -- (9.5,0) node[right] {0 MeV, 0$^+$, $T=0$};
\draw[very thick] (8,6.13) -- (9.5,6.13) node[right] {6.13 MeV, 3$^-$, $T=0$};
\draw[very thick] (8,7.12) -- (9.5,7.12) node[right] {7.12 MeV, 1$^-$, $T=0$};
\draw[very thick] (8,8.87) -- (9.5,8.87) node[right] {8.87 MeV, 2$^-$, $T=0$};
\draw[ultra thick] (8,12.97) -- (9.5,12.97) node[right] {12.97 MeV, 2$^-$, $T=1$};
\node[font=\Large] at (8.75,-0.4) {$^{16}$O};
\draw[->] (9.2,8.77) -- (9.2,7.22);
\draw[->] (9.2,7.02) -- (9.2,0.1);
\draw[->] (9,8.77) -- (9,6.23);
\draw[->] (9,6.03) -- (9,0.1);
% Draw the excitation levels for 16F
\draw[ultra thick] (10.5,15.42) -- (12,15.42);
\node[font=\Large] at (11.25,15) {$^{16}$F};
\node at (11.25,14.5) {(15.42 MeV)};
\draw[very thick] (10.5,15.61) -- (12,15.61);
\draw[very thick] (10.5,15.84) -- (12,15.84);
\node at (13.8,15.6) {0 MeV, 0$^-$, $T=1$};
\node at (14,16) {0.19 MeV, 1$^-$, $T=1$};
\node at (14,16.4) {0.42 MeV, 2$^-$, $T=1$};
% Draw the excitation levels for p+15O
\draw[->] (12.2,15.42) -- (13.3,14.88);
\draw[ultra thick] (13.5,14.88) -- (15,14.88);
\node[font=\Large] at (14.25,14.5) {p+$^{15}$O};
\node at (14.25,14) {(14.88 MeV)};
\end{tikzpicture}
\caption{Energy levels of $^{16}$N, $^{16}$O and $^{16}$F near the ground state with isospin $T$=1~\cite{ENSDF}.  
\label{fig:level}}
\end{figure}

The three states, $^{16}$N(g.s., 2$^-$), $^{16}$O(12.97 MeV, 2$^-$) and $^{16}$F(0.42 MeV, 2$^-$), form a $T=1$ triplet ($T_z=-1$, 0, 1). The energy levels of $^{16}$N, $^{16}$O and $^{16}$F near their ground states are shown in Fig.~\ref{fig:level}. Just above the ground state $^{16}$N(g.s., 2$^-$), there are also $T$=1 bound states at 0.120 MeV (0$^-$), 0.298 MeV (3$^-$) and 0.397 MeV (1$^-$). They all decay electromagnetically to $^{16}$N(g.s., 2$^-$), emitting a small $\gamma$ ray. We call these bound states including the ground state (2$^-$) as the g.s. group of $^{16}$N in the present report.  There are no bound states in $^{16}$F($T$=1).  

\section{Charged-current reaction $^{16}$O($\bar\nu_e, e^+$)$^{16}$N(g.s. group)}

The calculations of the electron/positron spectra from  $^{16}$O($\nu_e, e^-$)$^{16}$F and $^{16}$O($\bar\nu_e, e^+$)$^{16}$N reactions were fully described in Ref.~\cite{NakazatoCC} and the implications  of the $^{18}$O mixture in water on SN neutrino events were discussed in Ref~\cite{TSuzuki-CCO1618}. In this section, we discuss  on the CC reaction $^{16}$O($\bar{\nu}_e, e^+$)$^{16}$N(g.s. group), where the g.s. group consists of the bound states at 0 MeV (2$^-$), 0.120 MeV (0$^-$), 0.298 MeV (3$^-$) and 0.397 MeV (1$^-$). The bound states (0$^-$, 3$^-$, 1$^-$) decay electromagnetically to the ground state (2$^-$), emitting a small $\gamma$ ray. All of them are followed by the $\beta$ decay from $^{16}$N(g.s.) to $^{16}$O.
We discuss the g.s. group together, since the four bound states of the g.s. group in the CC reactions $^{16}$O($\bar{\nu}_e, e^+$)$^{16}$N(g.s. group) cannot be distinguished in a water-Cherenkov detector unless a small $\gamma$ ray (0.120 MeV, 0.298 MeV, 0.397 MeV) can be identified.  We describe some unique features of these CC reaction as compared to other CC reactions.
The formula of the cross section calculation for the  CC neutrino-oxygen reactions are given in Eq.(1) of Ref.~\cite{NakazatoCC} and we calculated the cross section of $^{16}$O($\bar{\nu}_e, e^+$)$^{16}$N(g.s. group) using the quenching factors $f_A$=0.68$\pm$0.05 and  $f_s$=0.65$\pm$0.05,  which were evaluated in the previous work~\cite{MSakuda, TSuzuki-CCO1618}. 
%%Since the transition strength of the multipole ($J$) is roughly proportional to $(2J +1)$, the contribution of the multipole (2$^-$) is the largest of the three multipoles. 

First, the reaction $\bar{\nu}_e + {\rm ^{16}O} \rightarrow e^+  + {\rm ^{16}N}$(g.s. group) can be uniquely identified by the coincidence 
 between a prompt positron from the primary reaction and a 6.13-MeV $\gamma$ ray (and partly 7.12-MeV and 8.87-MeV $\gamma$ rays) from the subsequent 
$\beta^-$ decay of  ${\rm ^{16}N}$(g.s.), ${\rm ^{16}N(g.s.)}  \rightarrow  {\rm ^{16}}$O$(E_x>0)+ e^- + \bar{\nu}_e $, both of which are produced at the same interaction point during the time interval of the $\beta$ decay. The detailed parameters of the $\beta$ decay of ${\rm ^{16}N(g.s.)}$ to ${\rm ^{16}O}$~\cite{ENSDF} are summarized in Table 1 and the schematic diagram of the decay is also shown in dashed lines in Fig.~\ref{fig:level}.  This coincidence method with the constraint on the same vertex position during on the decay interval will reduce the accidental background significantly. The identification of this reaction by applying the coincidence will reject other CC reactions $^{16}$O($\nu_e, e^-$)$^{16}$F and $^{16}$O($\bar\nu_e, e^+$)$^{16}$N($E_x>$1 MeV)~\cite{NakazatoCC}, which have larger cross sections than this reaction and have no delayed signals. Though some of these other CC reactions may be accompanied by the prompt $\gamma$ rays above 5 MeV which are emitted from the strong-decay products as $^{15}$N$^*$ or $^{15}$O$^*$ within a microsecond, they can be further removed if the second signals from the first one microsecond in the coincidence are excluded out of from the long decay time ($T_{1/2}$=7.13 sec).  

Secondly, the visible energy $T_{e^+}$ of the positron from the primary reaction of this channel (g.s. group) can be used to determine the incident neutrino energy as $E_{\bar{\nu}_e}=T_{e^+}+11.44$ MeV above the threshold energy ($E_{\rm th}$=11.44 MeV). A small $\gamma$ ray (0.12 MeV, 0.298 MeV, 0.397 MeV) of the g.s.  group is negligible as compared to 11.44 MeV. The electron or positron signal from other CC reactions cannot give the incident neutrino energy without knowing the excited states ($E_x$) of $^{16}$F and $^{16}$N.  This CC reaction from $^{16}$O to $^{16}$N(g.s. group) has the lowest energy threshold among the neutrino-oxygen reaction, except for the CC $^{18}$O($\nu_e, e^-$)$^{18}$F reaction ($E_{\rm th}$=1.66 MeV)~\cite{TSuzuki-CCO1618}. 

The cross section of $^{16}$O($\bar\nu_e, e^+$)$^{16}$N(g.s. group) is shown as a function of neutrino  energy in Fig.2 and also given in Table 2. 
The cross section of the CC reaction from $^{16}$O(g.s.) to $^{16}$N(g.s., 2$^-$) is the largest among the CC reaction to the g.s. group, that to the 1$^-$ state is about 3/5 of that to the 2$^-$ state below 50 MeV and that to the 0$^-$ state is about 15\% of that to the 2$^-$ state between 12 and 20 MeV. This feature is qualitatively explained by the strength of the transition ($S$=1 and $L$=1) proportional to ($2J+1$). Above 50 MeV, the cross section is contributed to by the transition ($S$=1 and $L$=3) from $^{16}$O(g.s.) to the 2$^-$ and 3$^-$ states. The cross section to the 2$^-$ state becomes even larger and that to the 3$^-$ state becomes significant, about 10\% of that to the 2$^-$ state, at higher energy than 100 MeV. 

The cross section of the CC reaction $^{16}$O($\bar\nu_e, e^+$)$^{16}$N(g.s. group) was first calculated by Haxton~\cite{Haxton}. We find that our calculation of the CC cross section for $^{16}$O($\bar\nu_e, e^+$)$^{16}$N(g.s. group) is larger by about 40\% than his calculation. We note that our evaluations of the  quenching factors for these  bound states, $f_s= 0.65\pm 0.05$ and $f_A= 0.68\pm 0.05$,  were validated by the transverse form factor $F^2_T(q)$ of the ($e, e^{\prime}$) cross section near 13 MeV (2$^{-}$, 1$^{-}$, 3$^{-}$)~\cite{Stroetzel, Kim, Sick}, the rate of the partial muon capture ($\mu^-, \nu_{\mu}$) from the 1s orbit on $^{16}$O(g.s., 0$^+$) to the bound states (2$^-$, 0$^-$, 3$^-$, 1$^-$) of $^{16}$N and the total muon capture rate from $^{16}$O to $^{16}$N(g.s., 2$^-$)~\cite{MSakuda}. 

 Below 30 MeV, this cross section of $^{16}$O($\bar\nu_e, e^+$)$^{16}$N(g.s. group) is dominant among all CC reactions. 
There are several excited states (2$^-$, 1$^-$) of $^{16}$N($E_x$=3-25 MeV) with significant CC cross sections and they all decay hadronically to $n+^{15}$N~\cite{NakazatoCC}. 
 Above 30 MeV, the cross section to the g.s. group becomes smaller than the sum of other CC cross sections by an order of magnitude. However, we note again that  the neutrino energy can be reconstructed, only when the electron/positron energy of other CC reactions can be measured and the excited states ($E_x$) of the nucleus in the reaction, either $^{16}$N or $^{16}$F, are measured for each event. In addition, those excited states ($E_x$) will decay hadronically and sometimes emit $\gamma$ rays from 5 to 9 MeV,  which will diffuse the primary electron/positron signal. On the other hand, the CC reaction $^{16}$O($\bar\nu_e, e^+$)$^{16}$N(g.s. group) can be unambiguously identified and the neutrino energy can be reconstructed, if the delayed coincidence method can be used. This is a unique feature. 

While the 6.13-MeV and 7.12-MeV states decay electromagnetically via $E3/E1$ transition to the ground state, producing a single 6.13-MeV and 7.12-MeV $\gamma$ ray, the 8.87-MeV state ($2^-$) decays mostly through cascade to the ground state, producing two $\gamma$ rays (2.74 MeV + 6.13 MeV, or 1.75 MeV + 7.12 MeV) and partly a single 8.87-MeV $\gamma$ ray. 
After the $\beta^-$ decay and the electromagnetic transitions, the branching ratios of emitting $\gamma$ rays becomes Br($E_{\gamma}$=6.13 MeV)=$0.662\pm 0.006$, Br(7.12 MeV)=0.048$\pm$0.004 and Br(8.87 MeV)=0.0106$\pm$0.0007~\cite{ENSDF} and the sum of them is Br($E_{\gamma}\geq $6.13 MeV)=$0.720\pm 0.007$. The probability of no  $\gamma$ ray is $0.280\pm 0.004$.  We note that the visible energy of the delayed signal is $E_d=E_{\gamma}+T_{\beta}=10.42\ {\rm MeV}-T_{\bar{\nu}_e}$ for all decay modes, including the decay of $^{16}$N(g.s) to $^{16}$O(g.s), where $T_{\bar{\nu}_e}$ is the neutrino energy from the $\beta^-$ decay,  ${\rm ^{16}N(g.s.)}  \rightarrow  {\rm ^{16}}$O$(E_x>0)+ e^- + \bar{\nu}_e $. The delayed signal $E_d$ is further contributed to  by the kinetic energy $T_{\beta}$ from the $\beta$ decay by 1-2 MeV as shown in Table 1 and this will make the detection efficiency larger. 

The $\beta$ decay of  $^{16}$N(g.s.)  goes  to  $^{16}$O(g.s.) with the branching ratio 28.0\%, producing an electron with $T_{{\beta} \max}=10.42$ MeV and no $\gamma$ ray. Even for this decay mode, the delayed coincidence between the primary positron and the delayed coincidence signal may be possible for the case of $T_{\beta} >$5 MeV. The kinetic energy spectrum of this $\beta$ decay mode is shown in Fig.3, where $Q$-value is equal to 10.42 MeV. About 49\% of the electrons have $T_{\beta}>$ 5 MeV and they can be detected. This will add a probability of about 14\% ($=0.28\cdot0.49$) to that of the delayed coincidence signal producing $\gamma$ rays with $E_{\gamma}\geq 6.13$ MeV (72\%) and the total probability of the delayed coincidence signal with visible energy greater than 5 MeV, from the $\beta$ decay of ${\rm ^{16}N(g.s.)}$, is estimated to be about 86\%.  

 The SNO experiment and SK experiment use $\gamma$ rays of 6.13 MeV and 7.12 MeV from $^{16}$N(g.s., 2$^-$) $\beta$ decay for a PMT calibration~\cite{SNOdet, SK-N16}.  The SK experiment also uses  a 6.13-MeV $\gamma$ ray to measure the NC neutrino-oxygen quasielastic interaction~\cite{T2K6MeV1, SK6MeV1, T2K6MeV2, SK6MeV2}, which are consistent with the calculation~\cite{Artur}. Thus, it is clear that a 6.13-MeV $\gamma$ ray can be observed in a water Cherenkov detector. 

We comment on the unique feature of this CC reaction channel to $^{16}$N(g.s. group). A detector such as SK-Gd~\cite{EGADS, SK-Gd1, Vagins} can measure the neutrino energy of the IBD reaction unambiguously by the neutron tagging and reconstruct the incident neutrino spectrum $F(E_{\bar{\nu}_e})$, using its well-known IBD cross section~\cite{Strumia, Vissani1} and the relation $E_{\bar{\nu}_e}=T_{e^+}$+1.80 MeV.   Then, we can measure the CC cross section $\sigma (E_{\bar{\nu}_e})$, using the measured neutrino spectrum $F(E_{\bar{\nu}_e})$, since we can measure $F(E_{\bar{\nu}_e})\cdot \sigma (E_{\bar{\nu}_e})$ from the measurement of the primary electron spectrum $T_{e^+}$, using the relation $E_{\bar{\nu}_e}=T_{e^+}$+11.44 MeV, where $\sigma (E_{\bar{\nu}_e})$ is the CC cross section to $^{16}$N(g.s. group). We can compare the measured cross section with our calculated one and confirm/improve the calculation. The measurement of this CC reaction will also validate the calculation of the NC cross sections which we describe in the next section, since the calculations of these CC and NC reactions are related by the CVC hypothesis~\cite{Feynman, Gerstein, MSakuda}. 
We also note that the delayed coincidence method to this reaction can be applied in the Hyper-K detector, even without the neutron tagging method.

\begin{table}[ht]
\caption{ Parameters of the $\beta^-$ decay of  $^{16}$N(g.s.), $^{16}$N(g.s.)$\rightarrow  {\rm  ^{16}O}(E_x)+ e^- + \bar{\nu}_e$~\cite{ENSDF}.  The half life of the decay is $T_{1/2}=7.13\pm 0.02$~sec. While the 6.13-MeV and 7.12-MeV states decay to the ground state, producing a single 6.13-MeV and 7.12-MeV $\gamma$ ray, the 8.87-MeV state ($2^-$) decays  through cascade to the ground state, producing mostly two $\gamma$ rays (2.74 MeV + 6.13 MeV, or 1.75 MeV + 7.12 MeV) and partly a single 8.87-MeV $\gamma$ ray.}
\centering
\begin{tabular}{c|ccc} 
 \hline
 $\beta^-$ decay of $^{16}$N(g.s.) to  &  Branching ratio  & $E_{\gamma}$  &   Range of $T_{\beta}$ \\
$E_x$ of $^{16}$O &  (\%)  & (MeV)  & (MeV) \\
 \hline
8.87 MeV  &  1.06$\pm$0.07 & 8.87 MeV & 0$<T_{\beta}<$1.55 MeV  \\
7.12 MeV  &  4.8$\pm$0.4 & 7.12 MeV & 0$<T_{\beta}<$3.30 MeV  \\
%% $\to$6.92 MeV  &  $<$0.34  & -  & $<$1  \\
6.13 MeV  &  66.2$\pm$0.6 & 6.13 MeV &  0$<T_{\beta}<$4.29 MeV   \\
0 MeV  & 28.0$\pm$0.4  & 0.0 & 0$<T_{\beta}<$10.42 MeV  \\

\hline
\end{tabular}
\end{table}

\begin{figure}[ht!]
\centering
\includegraphics[width=12.0cm,scale=1.0]{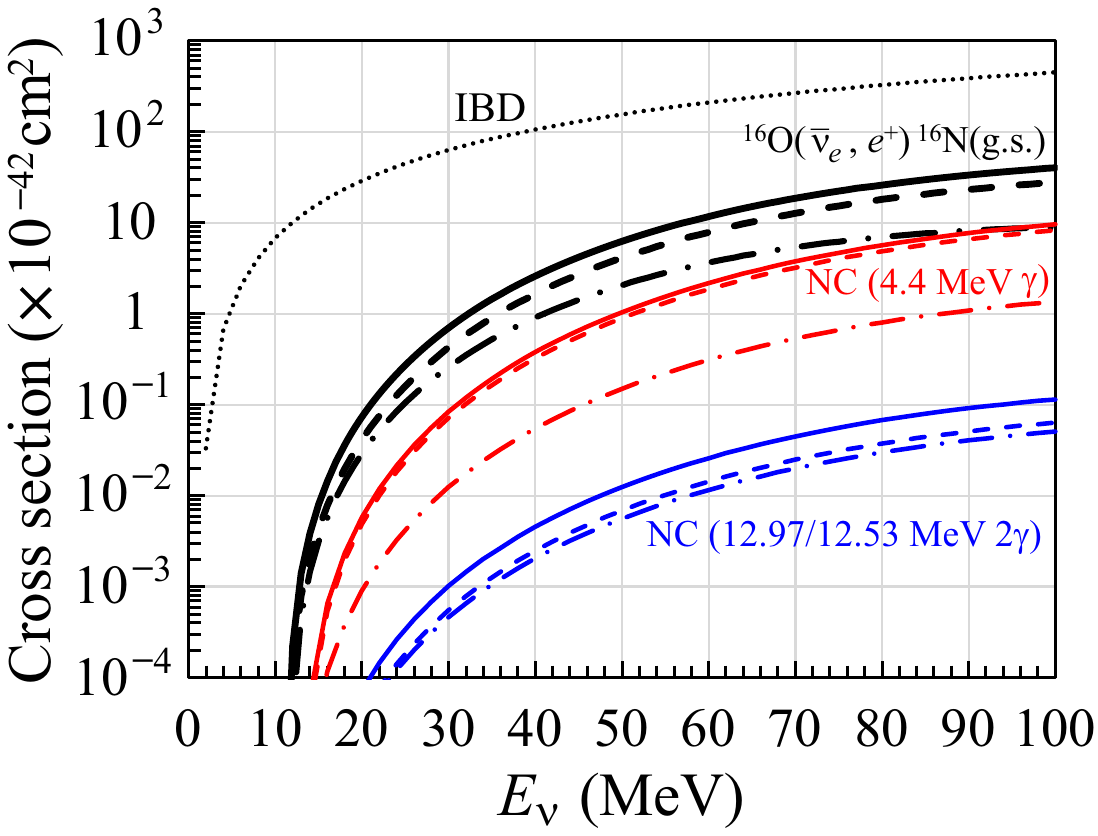}
\caption{The cross sections of the CC reactions $^{16}$O($\bar\nu_e, e^+$)$^{16}$N(g.s., $2^-$) (black dashed line), $^{16}$N($1^-$) (black dash-dotted line), and $^{16}$N(g.s. group) (black solid line); the NC cross sections of the 4.4-MeV $\gamma$ ray from $U$ and $D$ states, $\sigma^U_{\rm{NC},\gamma}$ (red dashed line) and $\sigma^D_{\rm{NC},\gamma}$ (red dash-dotted line), and the sum of them $\sigma^{tot}_{\rm{NC},\gamma}$ (red solid line);
the NC cross sections of the high-energy $\gamma$ rays (12.97 MeV and 12.53 MeV) via electromagnetic decay of the $U$ and $D$ states, $\sigma^U_{\rm{NC},2\gamma}$ (blue dashed line), $\sigma^D_{\rm{NC},2\gamma}$ (blue dash-dotted line), and the sum of them $\sigma^{tot}_{\rm{NC},2\gamma}$ (blue solid line)  as a function of the neutrino energy. The IBD cross section is shown in black dotted lines for comparison. 
}\label{fig:gamma}
\end{figure}

\begin{table}[t]
\caption{Cross sections of the CC reaction $^{16}$O($\bar{\nu}_e, e^+$)$^{16}$N(g.s. group) as functions of the neutrino energy, $E_\nu$ (MeV). The unit of the cross section is 10$^{-42}$ cm$^2$.  
\label{cslist}}
\centering\small
\begin{tabular}{c|ccccc}%%%The number of columns has to be defined here
\hline
%\multicolumn{1}{c|}{$E_\nu$} & \multicolumn{4}{c|}{$^{16}$O($\nu_e, e^-$)X} \\
$E_\nu$ (MeV) & 2$^-$ & 1$^-$ & 0$^-$ & 3$^-$ & Sum (2$^-$, 1$^-$, 0$^-$, 3$^-$)  \\
\hline
12	&1.59E$-$04	&4.04E$-$05	&2.51E$-$05	&0.0    &2.24E$-$04 \\
14  &2.10E$-$03	&1.46E$-$03	&3.03E$-$04	&0.0	&3.86E$-$03 \\
16  &7.79E$-$03	&5.74E$-$03	&8.69E$-$04	&0.0	&1.44E$-$02 \\
18	&1.98E$-$02	&1.44E$-$02	&1.73E$-$03	&1.20E$-$06	&3.60E$-$02 \\
20	&4.13E$-$02	&2.92E$-$02	&2.88E$-$03	&4.24E$-$06	&7.34E$-$02 \\
22	&7.57E$-$02	&5.20E$-$02	&4.32E$-$03	&1.23E$-$05	&1.32E$-$01 \\
24	&1.27E$-$01	&8.46E$-$02	&6.07E$-$03	&3.11E$-$05	&2.18E$-$01 \\
26	&2.00E$-$01	&1.29E$-$01	&8.13E$-$03	&7.07E$-$05	&3.37E$-$01 \\
28	&2.97E$-$01	&1.87E$-$01	&1.05E$-$02	&1.48E$-$04	&4.95E$-$01 \\
30	&4.25E$-$01	&2.60E$-$01 &1.32E$-$02	&2.88E$-$04	&6.99E$-$01 \\
32	&5.86E$-$01	&3.51E$-$01	&1.62E$-$02	&5.29E$-$04	&9.54E$-$01 \\
34	&7.86E$-$01	&4.60E$-$01	&1.95E$-$02	&9.27E$-$04	&1.27E+00 \\
36	&1.03E+00	&5.88E$-$01	&2.32E$-$02	&1.56E$-$03	&1.64E+00 \\
38	&1.31E+00	&7.36E$-$01	&2.72E$-$02	&2.52E$-$03	&2.08E+00 \\
40	&1.65E+00	&9.06E$-$01 &3.16E$-$02	&3.95E$-$03	&2.59E+00 \\
50	&4.11E+00	&2.06E+00	&5.90E$-$02	&2.57E$-$02	&6.25E+00 \\
60	&7.88E+00	&3.64E+00	&9.63E$-$02	&1.07E$-$01	&1.17E+01 \\
70	&1.27E+01	&5.37E+00	&1.44E$-$01	&3.31E$-$01	&1.85E+01 \\
80	&1.80E+01	&6.96E+00	&2.02E$-$01	&8.22E$-$01	&2.60E+01 \\
90	&2.32E+01	&8.18E+00	&2.69E$-$01	&1.73E+00	&3.34E+01 \\
100	&2.78E+01	&8.93E+00	&3.43E$-$01	&3.18E+00	&4.02E+01 \\
\hline
\end{tabular}
\end{table}%%%End of the table

\begin{figure}[ht]
\centering
\includegraphics[width=14.0cm,scale=1.0]{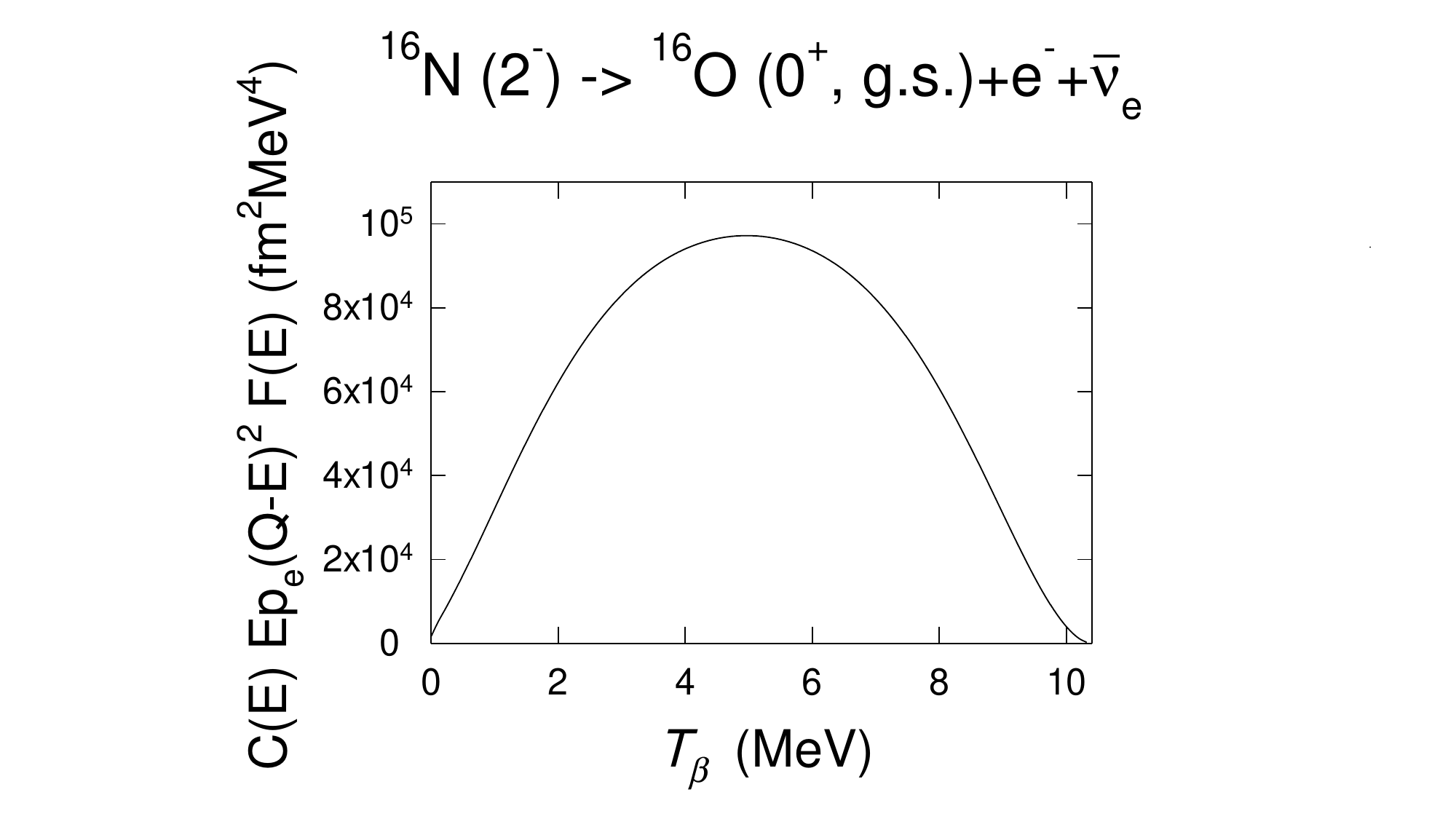}
\caption{
The electron energy spectrum of the $\beta$ decay, ${\rm ^{16}N(g.s.)}  \rightarrow  {\rm ^{16}}$O(g.s.)$+ e^- + \bar{\nu}_e $. The product of the nuclear shape factor $C(E)$, the phase space factor, and the Fermi function $F(E)$ is shown as a function of the electron kinetic energy $T_{\beta}$ (MeV). Here, $E$ and $p_e$ are electron energy and momentum, respectively, and $Q$ is the $Q$-value for the reaction.
\label{fig:beta}}
\end{figure}

\begin{table}[ht]
\caption{The two  $2^-$ states of $^{16}$O and their decay properties. The numbers are not yet established, but still uncertain~\cite{MSakuda}. We used ${\rm Br}^U(\alpha_1)$=0.35 and ${\rm Br}^D(\alpha_1)$=0.83. }\label{tab:neutrino}
\centering\small
\begin{tabular}{cccccc} 
 \hline
Excited states ($J^P,\ T$)  & $\Gamma$ & $\Gamma_{\alpha_1}$ &  $\Gamma_p$  & $\Gamma_{\gamma}$ &  Reference   \\
   &    (keV) &  (keV) &  (keV) &  (eV)  &      \\  
 \hline
12.53 MeV ($2^-, 0$)  &  &  &   &   &   \\
   & &   & &3.4$\pm$0.3 &   ~\cite{Gorodetzky}  \\
 &  0.097$\pm$0.010  & 0.072$\pm$0.010  & 0.025$\pm$0.003  & - &  ~\cite{Leavitt} \\
 &  0.108$\pm$0.010  & 0.092$\pm$0.010  & 0.016$\pm$0.003  & 0.55$\pm$0.06 &  ~\cite{Zijderhand}\\
%  &  0.111$\pm$0.010 & 0.072$\pm$0.10 & 0.025$\pm$0.03 & 3.4$\pm$0.3 & ENSDF~\cite{Tilley} \\
Values we used &  0.111$\pm$0.010 & 0.092$\pm$0.010   & 0.016$\pm$0.003 & 3.4$\pm$0.3   &  \\
\hline
 12.97 MeV ($2^-, 1$)  &    &  &  &   & \\
 &  &  0.69$\pm$0.07  & & 3.6$\pm$0.6 &   ~\cite{Gorodetzky}\\
 &  1.59$\pm$0.14  & 0.60$\pm$0.08  & 0.99$\pm$0.12  &  &  ~\cite{Leavitt} \\
  & 1.34$\pm$0.04 & 0.30$\pm$0.06  &1.04$\pm$0.07 & 1.6$\pm$0.3 & ~\cite{Zijderhand} \\
Values we used &  1.34$\pm$0.04 & &  &  3.6$\pm$0.6 & \\
\hline
\end{tabular}
\label{tab:nkmodel}
\end{table}

\section{Neutral-current reaction $^{16}$O($\nu, \nu^{\prime}$)$^{16}$O(12.53 MeV and 12.97 MeV, $2^-$) and the branching ratios of the two $2^-$ states producing $\gamma$ rays} 

We briefly review a feature of the NC reaction from the $U$ and $D$ states, which produces a 4.4-MeV $\gamma$ ray from the $\alpha$ decay of these states~\cite{MSakuda}. 
Next, we discuss a high-energy $\gamma$ ray emission from the electromagnetic decay of these two states. The latter cross sections for high-energy $\gamma$ rays are small, but the signature of the events is so distinct with high-energy visible energy above 10 MeV that those events can be clearly identified. 
In these calculations,  we use both the quenching factors of the  spin $g$ factor ($f_s=0.65$) and of the axial-vector coupling constant ($f_A=0.68$), and the isospin-mixing parameter of the two $2^-$ states  ($\beta=0.25$).
We summarize the decay properties of the two states in Table 3, which we use in the present paper as in the previous paper~\cite{MSakuda}. We also illustrate the NC reactions $^{16}$O($\nu, \nu^{\prime}$)$^{16}$O(12.53 MeV and 12.97 MeV, $2^-$) in Fig.4 which are relevant in this section.  

\begin{figure}[ht]
\centering
\includegraphics[width=12.0cm,scale=1.0]{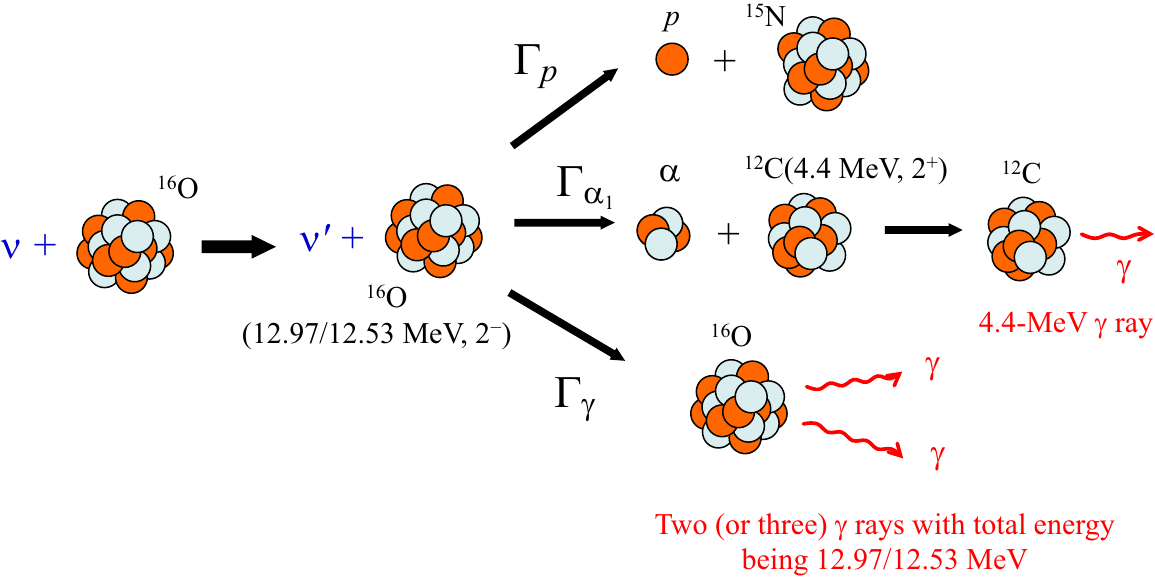}
\caption{Illustrative figure for the NC reactions, $ \nu  + {\rm ^{16}O}  \rightarrow  \nu   + {\rm ^{16}O(12.97/12.53 MeV,2^-)}$. \label{fig:NCillust}}
\end{figure}

\begin{table}[ht]
\caption{The branching ratios of the electromagnetic decay from the two states of $^{16}$O. Gorodetzky {\it et al.}~\cite{Gorodetzky} measured the widths of the cascade $\gamma$ rays as in the first column and we calculated the branching ratios in the second column to compare them with the measurements by Zijderhand and van der Leun~\cite{Zijderhand}. The sum of the branching ratios is normalized
to 100\% (the radiative width $\Gamma_{\gamma}$).}
\centering\small
\begin{tabular}{l|cc|c} 
 \hline
12.53 MeV ($2^-$) &   $\Gamma_{\gamma}$ (eV) & Branching ratio (\%) &  Branching ratio (\%) \\
Transition to the state &    &   &  \\
$\to$ 0 MeV ($0^+$) &  -  & - & 6.0$\pm$0.6  \\
$\to$ 8.87 MeV ($2^-$) &  0.86$\pm$0.10 & 25$\pm$3 & 33$\pm$2  \\
$\to$ 7.12 MeV ($1^-$) &  0.51$\pm$0.10 & 15$\pm$3 & 12.0$\pm$0.7  \\
$\to$ 6.13 MeV ($3^-$) &  2.1$\pm$0.2 & 60$\pm$6 &  49$\pm$2  \\
$\to$ All states  &  3.4$\pm$0.3  & 100 &  100 (0.55$\pm$0.06 eV) \\
Reference  & Gorodetzky {\it et al.}\cite{Gorodetzky}  &   & Zijderhand and van der Leun\cite{Zijderhand} \\
\hline
12.97 MeV ($2^-$) &   $\Gamma_{\gamma}$ (eV) & Branching ratio (\%) &  Branching ratio (\%) \\
Transition to the state &  &  &  \\
$\to$ 0 MeV ($0^+$) &  -  & -  & 2.1$\pm$0.4  \\
$\to$ 8.87 MeV ($2^-$) &  0.90$\pm$0.10 & 25$\pm$6 & 42$\pm$2  \\
$\to$ 7.12 MeV ($1^-$) &  0.44$\pm$0.10 & 12$\pm$2 & 6$\pm$1  \\
$\to$ 6.13 MeV ($3^-$) &  2.3$\pm$0.3  & 63$\pm$6  & 50$\pm$2  \\
$\to$ All states  &  3.6$\pm$0.3  & 100 & 100 (1.6$\pm$0.3 eV)  \\
Reference  & Gorodetzky {\it et al.}\cite{Gorodetzky} &   & Zijderhand and van der Leun\cite{Zijderhand} \\
\hline
\end{tabular}
\end{table}

First, we review the 4.4-MeV $\gamma$ ray from the $\alpha$ decay of the two states. The $\alpha$ decay of the 2$^-$ states to $\alpha +^{12}$C(0 MeV, 0$^+$) is forbidden by the angular momentum conservation and the $\alpha$ decay of the 2$^-$ states to $\alpha +^{12}$C(4.4 MeV, 2$^+$) is allowed through the $T$=0 component.  
The three experiments reported the $\alpha$-decay branching ratio Br($U\to \alpha + ^{12}$C(4.4 MeV))= $\Gamma_{\alpha_1}/\Gamma$, which we denote as  Br$^U(\alpha_1)$: Leavitt~${\it et\ al.}$~\cite{Leavitt}, Zijderhand and van der Leun~\cite{Zijderhand}, and Charity ${\it et\ al.}$~\cite{Charity} reported Br$^{U}(\alpha_1)$ to be 0.37$\pm$0.06, 0.22$\pm$0.04, and  0.46$\pm$0.08, respectively. We took a simple mean of the three values~\cite{Leavitt, Zijderhand, Charity} and used this mean value, ${\rm Br}^{U}(\alpha_1)$=0.35, to evaluate the 4.4-MeV $\gamma$-ray production cross section in the previous paper~\cite{MSakuda} as well as in the present paper.
We denote the NC cross section of the $U$ and $D$ states  as  $\sigma^U_{\rm NC}$ and  $\sigma^D_{\rm NC}$, respectively, and  the sum of them as  $\sigma^{\rm tot}_{\rm NC}=\sigma^U_{\rm NC} +\sigma^D_{\rm NC}$. We also denote the 4.4-MeV $\gamma$-ray production cross section of the $U$ and $D$ states  as  $\sigma^U_{{\rm NC},\gamma}$ and $\sigma^D_{{\rm NC},\gamma}$, respectively, and  the sum of them as  $\sigma^{\rm tot}_{{\rm NC},\gamma}=\sigma^U_{{\rm NC},\gamma} +\sigma^D_{{\rm NC},\gamma}$. 
We note that the NC cross sections  $\sigma^U_{\rm NC}$ and  $\sigma^D_{\rm NC}$ are calculated for  an average of one neutrino flavor  and its anti-neutrino flavor. 
In the previous work~\cite{MSakuda}, only the figures for the $U$ state, $\sigma^U_{\rm  NC}$ and $\sigma^U_{{\rm NC}, \gamma}=\sigma^U_{\rm NC}\cdot {\rm Br}^U(\alpha_1)$, were shown. This time, $\sigma^U_{{\rm NC},\gamma}$ (red dashed line) and  $\sigma^D_{{\rm NC},\gamma}= \sigma^D_{\rm NC}\cdot {\rm Br}^D(\alpha_1)$ (red dash-dotted line), with ${\rm Br}^U(\alpha_1)$=0.35 and ${\rm Br}^D(\alpha_1)$=0.83, respectively, are shown in Fig.2. The sum of them $\sigma^{\rm tot}_{{\rm NC},\gamma}$  (red solid line) is also shown and it is larger than $\sigma^U_{{\rm NC},\gamma}$ by about 16\%, since the ratio $\sigma^D_{{\rm NC},\gamma}/\sigma^U_{{\rm NC},\gamma}$=0.16 at $\beta$=0.25.

Next, we discuss the high-energy $\gamma$-ray production via electromagnetic decay of the two  states. The direct $M2$ electromagnetic transition from the two $2^-$ states to the ground state $0^+$ is suppressed and electromagnetic transitions to the ground state go through the cascade transitions producing more than two $\gamma$ rays. Gorodetzky ${\it et\ al.}$~\cite{Gorodetzky} measured both the electromagnetic cascade decay $^{15}$N($p, \gamma \gamma$)$^{16}$O and the $\alpha$ decay $^{15}$N($p, \alpha_1 \gamma) ^{12}$C(4.4 MeV) from the $U$ and $D$ states in the proton capture experiment. Zijderhand and van der Leun~\cite{Zijderhand} also measured both  a single  $\gamma$ ray from the electromagnetic decay $^{15}$N($p, \gamma$)$^{16}$O and the $\alpha$ decay $^{15}$N($p, \alpha_1 \gamma) ^{12}$C(4.4 MeV) from the two states in the proton capture experiment. Their values are shown in Table 3. We use the radiative decay widths measured by Gorodetzky ${\it et\ al.}$~\cite{Gorodetzky} in the present paper. The latest evaluation for the decay parameters of $^{16}$O can be found in Ref.~\cite{ENSDF} and the values for the radiative decay widths are the same as those we use in the present paper. As shown in Table 3, we use the total decay widths $\Gamma$=0.111$\pm$0.010 (keV) and $\Gamma$=1.34$\pm$0.04 (keV) for the $D$ and $U$ states, respectively, to calculate the branching ratios of the two states producing $\gamma$ rays, Br($D$ $\to$ $\gamma$ rays)=$\Gamma_{\gamma}/\Gamma$=3.1$\pm$0.04\% and Br($U$ $\to$ $\gamma$ rays)=0.27$\pm$0.02\%. 
%%~\footnote{ENSDF gives 3.2$\pm$0.04\% and 0.28$\pm$0.02\% for $\Gamma_{\gamma}/\Gamma$ of the $D$ and $U$ states, respectively. We used $\Gamma_{\gamma}/\Gamma$=3.1$\pm$0.04\% and 0.27$\pm$0.02\%, respectively, for consistency in this paper.}. 
If we take $\Gamma_{\gamma}$=3.4$\pm$0.3 eV of Gorodetzky ${\it et\ al.}$~\cite{Gorodetzky} for the $D$ state, we naturally obtain the total width $\Gamma$ to be 
$\Gamma =\Gamma_{\alpha}+\Gamma_p +\Gamma_{\gamma}$=0.092+0.016+0.0034=0.1114 (keV), which agrees with the total width of the $D$ state listed in Ref.~\cite{ENSDF}. Thus, the branching ratio of the $D$ state producing  $\gamma$ rays is $\Gamma_{\gamma}/\Gamma$=3.1$\pm$0.04\%. We note that Zijderhand and van der Leun~\cite{Zijderhand} adopted the total width $\Gamma$=0.108 keV by taking the average of the three previous measurements of the total width~\cite{Maurel, Damjan, Leavitt}, including the measurement by Leavitt ${\it et\ al.}$~\cite{Leavitt}. Thus, the measurements of the branching ratios for the $D$ state are consistent with each other except for the $\gamma$-ray width.  On the other hand, the measurements for the $U$ states are not consistent with each other. 

As shown in Table 4, the cascade electromagnetic decay of the $U$  state produces a pair of two (sometimes three) $\gamma$ rays, 6.84 MeV + 6.13 MeV (63\%), 5.85 MeV + 7.12 MeV (12\%), 4.10 MeV + 8.87 MeV (25\%). As explained in the previous section, the 8.87-MeV state decays mostly through cascade to the ground state, producing two $\gamma$ rays. Similarly, the cascade electromagnetic decay of the $D$  state produces a pair of two (sometimes three) $\gamma$ rays, 6.40 MeV + 6.13 MeV (60\%), 5.41 MeV + 7.12 MeV (15\%), 3.66 MeV + 8.87 MeV (25\%).  We note that the electromagnetic decay will produce mostly two  $\gamma$ rays at the same time and at the same vertex position and that the sum of them is as high as 12.97 MeV and 12.53 MeV from the $U$ and $D$ states, respectively. This feature can be used to identify this reaction channel and the coincidence method can be also used if the multiple-ring reconstruction can be developed~\cite{SK-tau}. 

We show  the cross section of the high-energy $\gamma$-ray production via electromagnetic decay of the $U$ and $D$ states in Fig.2 and also in Table 5. We denote them as $\sigma^U_{\rm{NC},2\gamma}= \sigma^U_{\rm{NC}}\cdot {\rm Br}(U \to \gamma$ rays), $\sigma^D_{\rm{NC},2\gamma}= \sigma^D_{\rm NC}\cdot {\rm Br}(D \to \gamma$ rays) and the sum of them $\sigma^{\rm tot}_{{\rm NC},2\gamma}=\sigma^U_{{\rm NC},2\gamma} +\sigma^D_{{\rm NC},2\gamma}$.  Since the branching ratios for the $U$ and $D$ states are ${\rm Br}(U \to \gamma \ {\rm rays})= 0.27\pm 0.02$\% and ${\rm Br}(D \to \gamma \ {\rm rays})=3.1\pm 0.04$\%, respectively, each of their cross sections is smaller than the CC cross section by three orders of magnitude. However, it should be reminded  that 
all flavors of neutrinos and antineutrinos contribute to the number of NC events while the CC reaction depends on the flavor.
 %We adopt the values by The measurement of the radiative  is also uncertai and n. 
 Only new measurements will resolve the above  inconsistencies in the branching ratios between Gorodetzky ${\it et\ al.}$~\cite{Gorodetzky} and Zijderhand and van der Leun~\cite{Zijderhand}.

\begin{table}[t]
\caption{NC cross sections of 4.4-MeV $\gamma$-ray production ($\sigma^U_{{\rm NC},\gamma}$  and $\sigma^D_{{\rm NC},\gamma}$), 12.97-MeV and 12.53-MeV $\gamma$-ray production ($\sigma^U_{{\rm NC},2\gamma}$ and $\sigma^D_{{\rm NC},2\gamma}$) from the $U$ and $D$ states as functions of the neutrino energy, $E_\nu$ (MeV). We note that each of the NC cross section is calculated for an average over one neutrino flavor and its anti-neutrino flavor. The unit of the cross section is 10$^{-42}$ cm$^2$.  
\label{cslist1}}
\centering\small
\begin{tabular}{c|cccc}%%%The number of columns has to be defined here
\hline
%\multicolumn{1}{c|}{$E_\nu$} & \multicolumn{4}{c|}{$^{16}$O($\nu_e, e^-$)X} \\
$E_\nu$ (MeV)&  $\sigma^U_{{\rm NC},\gamma}$ & $\sigma^D_{{\rm NC},\gamma}$  & $\sigma^U_{{\rm NC},2\gamma}$  & $\sigma^D_{{\rm NC},2\gamma}$   \\
\hline
12 &0 & 0	& 0	& 0 \\
14 	& 4.61E$-$05	& 1.55E$-$05		& 3.56E$-$07	& 5.81E$-$07 \\
16 	& 5.41E$-$04	& 1.17E$-$04		& 4.17E$-$06	& 4.37E$-$06 \\
18 	& 1.95E$-$03	& 3.82E$-$04		& 1.51E$-$05	& 1.43E$-$05 \\
20 	& 4.88E$-$03	& 9.10E$-$04		& 3.76E$-$05	& 3.40E$-$05 \\
22 	& 1.00E$-$02	& 1.82E$-$03		& 7.74E$-$05	& 6.81E$-$05 \\
24 	& 1.83E$-$02	& 3.26E$-$03		& 1.41E$-$04	& 1.22E$-$04 \\
26 	& 3.05E$-$02	& 5.39E$-$03		& 2.36E$-$04	& 2.01E$-$04 \\
28 	& 4.79E$-$02	& 8.37E$-$03		& 3.69E$-$04	& 3.13E$-$04 \\
30 	& 7.14E$-$02	& 1.24E$-$02		& 5.51E$-$04	& 4.63E$-$04 \\
32 	& 1.02E$-$01	& 1.76E$-$02		& 7.89E$-$04	& 6.59E$-$04 \\
34 	& 1.42E$-$01	& 2.43E$-$02		& 1.09E$-$03	& 9.08E$-$04 \\
36 	& 1.91E$-$01	& 3.26E$-$02		& 1.47E$-$03	& 1.22E$-$03 \\
38 	& 2.50E$-$01	& 4.26E$-$02		& 1.93E$-$03	& 1.59E$-$03 \\
40 	& 3.22E$-$01	& 5.47E$-$02		& 2.48E$-$03	& 2.04E$-$03 \\
50 	& 8.90E$-$01	& 1.49E$-$01		& 6.87E$-$03	& 5.58E$-$03 \\
60 	& 1.86E+00	& 3.10E$-$01		& 1.43E$-$02	& 1.16E$-$02 \\
70 	& 3.21E+00	& 5.33E$-$01		& 2.48E$-$02	& 1.99E$-$02 \\
80 	& 4.85E+00	& 8.02E$-$01		& 3.74E$-$02	& 3.00E$-$02 \\
90 	& 6.59E+00	& 1.09E+00		& 5.08E$-$02	& 4.06E$-$02 \\
100 & 8.24E+00	& 1.36E+00	& 6.35E$-$02	& 5.07E$-$02 \\
\hline
\end{tabular}
\end{table}%%%End of the table

%%%%%%%%%%%%%%%%%%%%%%%%%%%%%%%%%%%%%%%%%%%%%\fi
\section{Estimation of the number of events from  $^{16}$O($\bar\nu_e, e^+$)$^{16}$N(0 MeV, $2^-$) and $^{16}$O($\nu, \nu^{\prime}$)$^{16}$O(12.97/12.53 MeV, $2^-$), induced by supernova neutrinos}\label{sec5}

We evaluate the number of these CC and NC events induced by neutrinos from SN explosion which can be observed by the SK, a 32 kton water Cherenkov detector~\cite{SK-det} in the Earth~\footnote{The evaluations in this section can be applied to the Hyper-K detector (187-kton fiducial volume)~\cite{HyperK-SN} if the detector threshold on the electron kinetic energy is taken into account. The SK detector can measure the kinetic energy of electron/positron larger than 3.5 MeV~\cite{SK-det} and the recent Hyper-K study on SN uses 5 MeV for the threshold on the kinetic energy.}. The four bound states of the g.s. group of $^{16}$N, 0 MeV (2$^-$), 0.120 MeV (0$^-$), 0.298 MeV (3$^-$) and 0.397 MeV (1$^-$), in the CC reactions cannot be distinguished  in a water-Cherenkov detector unless a small $\gamma$ ray  can be identified. We thus estimate the number of $^{16}$O($\bar\nu_e, e^+$)$^{16}$N(0 MeV, $2^-$) and $^{16}$N(g.s. group). 

We calculate the number of events using the following parametrization (called KRJ fit~\cite{Keil, Tamborra}) for the normalized SN neutrino spectra $f(E_\nu)$ as we used in the previous work~\cite{MSakuda}:
\begin{equation}
f(E_\nu) =  \frac{(\alpha+1)^{\alpha+1}}{\Gamma(\alpha+1) \langle E_\nu\rangle ^{\alpha+1}}E_\nu^\alpha\exp\Big{(}-\frac{(\alpha+1)E_\nu}{\langle E_\nu\rangle} \Big{)},
\label{eq:krj}
\end{equation}
where $\langle E_\nu\rangle$ is the average neutrino energy. In this expression, $\Gamma(\alpha+1)$ is the Gamma function and $\alpha$ is the pinching parameter.  As the value $\alpha $ becomes larger, the high-energy tail of the distribution is more suppressed for the same average energy. We only calculate the number of events using three typical values of the KRJ fit  with $\alpha=3$ and $\langle E_\nu\rangle=$10, 12 and 14  MeV,  and assume that the neutrino spectra are flavor independent in Table 6.

%%\textcolor{green}{
The time-integrated number spectrum of neutrinos emitted from a SN core, $dN_{\nu}/dE_{\nu}$, is related to the normalized neutrino spectra $f(E_\nu)$ as
\begin{equation}
    \frac{dN_{\nu}}{dE_{\nu}} = \frac{E_\nu^{\rm tot}}{\langle E_\nu\rangle} f(E_\nu),
    \label{eq:tintspec}
\end{equation}
where $E_\nu^{\rm tot}$ is the total energy emitted by one neutrino flavor. Hereafter, we set $E_\nu^{\rm tot}=5\times 10^{52}$~erg for each neutrino flavor. We calculate the number of events at various average energies using the neutrino flux $F(E_\nu)$ at a detector on the Earth, which is given as
\begin{equation}
    F(E_\nu) = \frac{1}{4\pi d_{\rm SN}^2}\frac{E_\nu^{\rm tot}}{\langle E_\nu\rangle} f(E_\nu).
    \label{eq:flux}
\end{equation}
We set the distance from a detector to the SN to $d_{\rm SN}=10$~kpc.

\begin{figure}[ht!]
\centering
\includegraphics[width=10.0cm,scale=1.0]{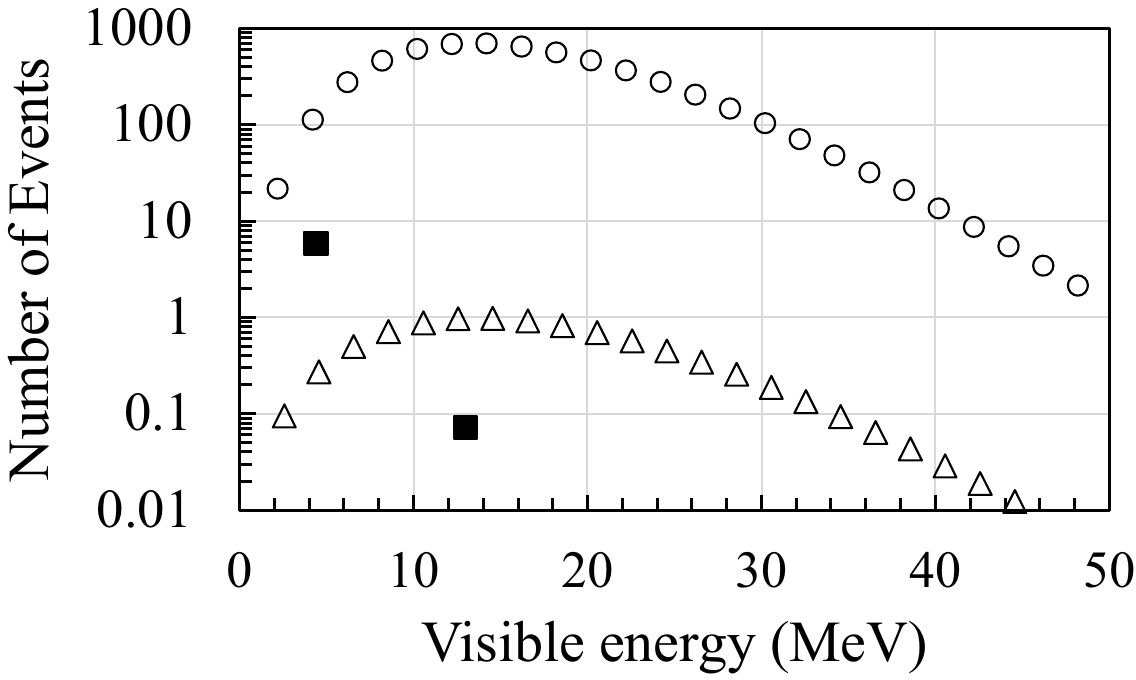}
\caption{The visible energy spectrum of $^{16}$O($\bar\nu_e, e^+$)$^{16}$N(g.s. group) (open triangles) and that of  the IBD event spectrum (open circles) are also plotted with an energy-bin width of 2 MeV as a function of the visible energy $E_{\rm vis}$ for the KRJ fit with $\alpha=3$ and $\langle E_\nu\rangle=12$ MeV.  The total numbers of those NC $\gamma$ events due to  $\sigma^{\rm tot}_{{\rm NC}, \gamma}$  and $\sigma^{\rm tot}_{{\rm NC}, 2\gamma}$ are plotted in filled black squares at $E_{\rm vis} = 4.4$ MeV and at $E_{\rm vis} = 12.97$ MeV, for the same KRJ fit values. 
\label{fig:SKspectra}}
\end{figure}

We calculate the number of events $N^{(i)}(E_{\nu})$ produced in the energy range from $E_{\nu}$ to $E_{\nu}+\Delta E_{\nu}$ for the IBD reaction and the CC reactions $^{16}$O($\bar\nu_e, e^+$)$^{16}$N(g.s. group) by folding the neutrino flux and the cross sections as,
\begin{eqnarray}
N^{(i)}(E_{\nu}) = n_{\rm tar}^{(i)} F(E_{\nu})\sigma ^{(i)}(E_{\nu})\Delta E_{\nu},
\label{eq:cc}
\end{eqnarray}
where $\sigma ^{(i)}(E_{\nu})$ stands for the cross section of either IBD or CC reaction and $n_{\rm tar}^{(i)}$ is the number of either protons or $\rm^{16}O$ targets for the case of a 32 kton fiducial volume of the SK  detector~\cite{SK-SN}. 
For the IBD reaction (CC reactions), the relation $E_{\nu}=T_{e^+}+1.80$ MeV (11.44 MeV) between the neutrino energy $E_{\nu}$ and the visible energy $E_{\rm vis}$ ($=T_{e^+}$) of the positron holds. 
In Fig.~\ref{fig:SKspectra}, the number of events of the  CC reactions are compared with that of the IBD reaction as a function of the visible energy $E_{\rm vis}$ with an energy width of $\Delta E_{\nu}=2$ MeV, for the KRJ fit with $\alpha=3$ and $\langle E_\nu\rangle=12$ MeV. The numbers of events of the CC reactions integrated over the neutrino energy up to 100 MeV are summarized in Table 6, for the KRJ fit  with $\alpha=3$ and $\langle E_\nu\rangle=10$, 12 and 14  MeV, where we show the effect of the requirement on the kinetic energy of a positron with $T_e>5$ MeV or $T_e>0$ MeV (Threshold). The primary positron spectrum from the CC reaction depends on the assumed SN flux and the cross section, and about 2\% of the spectrum lies below 5 MeV for the KRJ fit ($\alpha=3$ and $\langle E_{\nu}\rangle =12$ MeV). 
We do not consider the efficiency of having the coincidence signal with the visible energy greater than 5 MeV, which is estimated to be about 86\%. We point out again the importance of the low threshold energy of the CC reactions, since 98\%, 54\% and only 18\% of the SN neutrino flux $F(E_\nu)$ remain after the requirement of $E_{\nu}>1.80$ MeV (IBD reaction), 11.44 MeV (CC $^{16}$N(g.s.) reaction) and 18 MeV (typical $E_{\rm th}$ value for CC $^{16}$N($E_x> 3$ MeV) reactions)~\cite{NakazatoCC}, respectively, for the typical KRJ fit with $\alpha=3$ and $\langle E_\nu\rangle=12$ MeV, and the effect of the requirement on the positron kinetic energy $T_e$ will be imposed additionally.

For the NC reactions, the $\gamma$ ray of 4.4 MeV, 12.53 MeV or 12.97 MeV is produced independently of the incident neutrino energy $E_{\nu}$, we can  calculate only the total number of events $N^{(i)}$ integrated over $E_\nu$ as    
\begin{eqnarray}
N^{(i)} = n_{\rm tar} \int_0^{E_\nu^{\rm max}} dE_\nu F(E_\nu) \sigma^{(i)}(E_{\nu}) ,
\label{eq:nc-gamma}
\end{eqnarray}
where $\sigma ^{(i)}(E_{\nu})$ stands for the cross section of any type of NC reactions and  $n_{\rm tar}$ is the number of $\rm^{16}O$ targets in a 32 kton fiducial volume of the SK  detector and we set $E_\nu^{\rm max}=100$ MeV.

We show in Table 6 the total numbers of the NC events containing 4.4-MeV $\gamma$ rays due to $\sigma^{U}_{{\rm NC},\gamma}$  and $\sigma^{\rm tot}_{{\rm NC},\gamma}$ as well as  those containing  12.97-MeV and 12.53-MeV $\gamma$ rays due to $\sigma^{U}_{{\rm NC},2\gamma}$ and $\sigma^{\rm tot}_{{\rm NC},2\gamma}$, for the KRJ fit with $\alpha=3$ and $\langle E_\nu\rangle=$10, 12 and 14  MeV. We also plot in Fig.5 the total numbers of those NC $\gamma$ events due to $\sigma^{\rm tot}_{{\rm NC}, \gamma}$ at $E_{\rm vis} = 4.4$ MeV and $\sigma^{\rm tot}_{{\rm NC}, 2\gamma}$ at $E_{\rm vis} = 12.97$ MeV in filled black squares, for the KRJ fit with $\alpha=3$ and $\langle E_\nu\rangle=12$ MeV, assuming that the detection efficiency is 100\%  without considering the energy resolution.  

The number of events due to the $^{16}$O($\bar\nu_e, e^+$)$^{16}$N(g.s. group) reaction is smaller by two or three orders of magnitude than that of the IBD events. The total cross section of NC events are summed over 3 flavors of a neutrino  and anti-neutrino, namely, 6 times the average NC cross section. That explains why the number of the  NC events producing a 4.4-MeV $\gamma$ ray  due to $\sigma^{U}_{{\rm NC},\gamma}$  and $\sigma^{\rm tot}_{{\rm NC},\gamma}$ is nearly the same as that of the $^{16}$O($\bar\nu_e, e^+$)$^{16}$N(g.s. group) reaction.
The numbers of NC events producing 12.97 and 12.97/12.53-MeV $\gamma$ ray due to $\sigma^{U}_{{\rm NC},2\gamma}$  and $\sigma^{\rm tot}_{{\rm NC},2\gamma}$, respectively, are smaller by two orders of magnitude than that of the $^{16}$O($\bar\nu_e, e^+$)$^{16}$N(g.s. group) reaction.

\begin{table}[h]
\caption{Expected number of neutrino events from a core-collapse SN at 10 kpc to be detected at SK (32-kton fiducial volume) for the models with KRJ fit.}
\centering\small
\begin{tabular}{lllccc} 
\hline
 \hline
& KRJ parameters ($\alpha,  \langle E_{\nu} \rangle$ [MeV])& Condition  &  (3, 10)  &  (3, 12) & (3, 14)  \\
 \hline
IBD & $p(\bar{\nu}_e,e^+)n$ & $T_e>$0 MeV & 4.83$\times$10$^3$  & 4.88$\times$10$^3$ & 6.88$\times$10$^3$ \\
& & $T_e>$5.0 MeV & 4.71$\times$10$^3$  & 4.81$\times$10$^3$ & 6.84$\times$10$^3$ \\
 \hline
CC & $^{16}$O($\bar\nu_e, e^+$)$^{16}$N(g.s., $2^-$) & $T_e>$0 MeV
 & 1.6 & 4.6 & 10.6 \\
& & $T_e>$5.0 MeV  & 1.5 & 4.5 & 10.5 \\
& $^{16}$O($\bar\nu_e, e^+$)$^{16}$N(g.s. group) & $T_e>$0 MeV 
 & 2.8 & 7.8 & 17.4 \\
& & $T_e>$5.0 MeV  & 2.6 & 7.6 & 17.2 \\
 \hline
% NC $\rm^{16}O(\nu,\nu^\prime)\rm^{16}O^*(12.97MeV)$, %$E_{\gamma}=$4.4 MeV) &  1.9 & 5.1 & 10.7 & 13 & 158 \\
NC & $\sigma^U_{{\rm NC},\gamma}$ & $E_{\gamma}=4.4$ MeV &  1.9 & 5.1 & 10.7 \\
&  $\sigma^{\rm tot}_{{\rm NC},\gamma}$ & $E_{\gamma}=4.4$ MeV &  2.2 & 5.9 & 12.4 \\
& $\sigma^U_{{\rm NC},2\gamma}$ & $E_{2\gamma}$=12.97 MeV &  0.015 & 0.039 & 0.082 \\
& $\sigma^{\rm tot}_{{\rm NC},2\gamma}$ & $E_{2\gamma}$=12.97 or 12.53 MeV &  0.027 & 0.072 & 0.151 \\
 \hline
 \hline
\end{tabular}
\label{tab:evnmb1}
\end{table}

\section{Summary}\label{sec6}
%\textcolor{red}{Sakuda, 2022.1031: 
In the previous work, we discussed the detection of a single 4.4-MeV $\gamma$ ray produced in the neutrino NC reaction $^{16}$O($\nu, \nu^{\prime}$)$^{16}$O(12.97 MeV and 12.53 MeV, $2^-$), after evaluating both the isospin-mixing parameter $\beta$ and the quenching factors, $f_s=g_s^{{\rm eff}}/g_s$ of the  spin $g$ factor and $f_A=g_A^{{\rm eff}}/g_A$ of the axial-vector coupling constant $g_A$, of the two $2^-$ states~\cite{MSakuda}. 
In this report, we have further examined both  the $\beta$ decay of the CC reaction $^{16}$O($\bar{\nu}_e$, e$^{+}$)$^{16}$N(g.s. group) and  the electromagnetic decay of the NC reaction $^{16}$O($\nu, \nu^{\prime}$)$^{16}$O(12.97 MeV and 12.53 MeV, $2^-$) producing high-energy $\gamma$ rays of 12.97 MeV and 12.53 MeV, using the same quenching factors for the two $2^-$ states.  We have evaluated the number of these CC and NC reactions induced by neutrinos from SN explosion which can be observed by the SK, a 32 kton water Cherenkov detector~\cite{SK-det} in the Earth. 
 
Even though the cross sections of these CC/NC reactions are small, the application of the present work for the future SN neutrino events by water Cherenkov detectors has some unique features. Above all,  the coincidence method can be applied for the more solid identification in the  former CC reaction.  In the latter NC reaction, the large visible energy of 12.97 MeV and 12.53 MeV can be produced and the coincidence method can be also used  if the multiple-ring reconstruction can be developed~\cite{SK-tau}. Further, it is important for the study of SN physics to have the detection channels which have significant sensitivities at as low neutrino energy as 10 to 20 MeV, where the majority of the neutrino energy spectra from SN bursts lie. These neutrino-oxygen reactions have the lowest energy thresholds (11.44 MeV and 12.97 MeV)  among the neutrino-oxygen reactions, which correspond to the energy levels of the $T$=1 ground states of $^{16}$N and $^{16}$O.  They have the dominant cross sections from 12 to 30 MeV, above which the cross sections of other CC and NC reactions from the $T$=1 excited states dominate. 

We also note that the delayed coincidence method to this reaction can be applied in the Hyper-K detector, even without the neutron tagging method using Gd.  The recent study by the Hyper-K Collaboration on the  detection of the SN neutrino events~\cite{HyperK-SN} comments that they do not consider the $\gamma$-ray emission from the NC interactions on $^{16}$O nuclei, since a dominant channel $^{16}$O($\nu, \nu')^{16}$O($E_x>$16 MeV) mainly produces only $\gamma$ rays with an energy of 5 MeV to 9 MeV~\cite{Langanke} and the visible energy from these events would typically be below 5 MeV (Hyper-K energy threshold) after Compton scattering on an electron or electron-positron pair production.   
Our study for the coincidence method including the high energy $\gamma$ rays may turn out to be useful. 

The JUNO experiment~\cite{JUNO} uses the NC reaction $^{12}$C($\nu, \nu^{\prime}$)$^{12}$C(15.11 MeV, $1^+$) and CC reactions $^{12}$C($\nu_e, e^-$)$^{12}$N(g.s., $1^+$) with the subsequent $\beta$ decay and $^{12}$C($\bar{\nu}_e, e^+$)$^{12}$B(g.s., $1^+$) with the subsequent $\beta$ decay~\cite{Fukugita}, as the main detection channels for the analysis of the SN neutrino bursts, in addition to the IBD reaction, elastic $\nu$–p scattering and elastic  $\nu$–e scattering~\cite{Strumia, Vissani1, Beacom}. It is important to note some basic features of the CC/NC neutrino-oxygen reactions which are different from those of the CC/NC neutrino-carbon reactions.  
The neutrino-carbon cross sections related to the $1^+$ state are larger by two orders of magnitude than the neutrino-oxygen cross sections from the 12.97 and 12.53 states ($2^-$) since the former neutrino-$^{12}$C reactions have a large matrix element causing the spin-flip transition from $1p_{3/2}$ to $1p_{1/2}$, while the latter neutrino-$^{16}$O reactions go through the spin-dipole transition from fully occupied 1$p$ shells to $2s-1d$ shells, which is smaller by an order of magnitude than the former.  
In addition, the electromagnetic decay branching ratio ($\Gamma_{\gamma}/\Gamma$=96\%) of the former state $^{12}$C(15.11 MeV, $1^+$)~\cite{Kelley} is larger by two order of magnitude  than that ($\Gamma_{\gamma}/\Gamma$) of the $2^-$ states of $^{16}$O. This is because in the former state $^{12}$C(15.11 MeV), the electromagnetic decay is dominant and the hadronic decay to $p+^{11}$B decay is suppressed due to the threshold ($E_{\rm th}=15.96$ MeV), while in the latter $2^-$ states of $^{16}$O,  the hadronic decays are allowed ($E_{\rm th}=12.13$ MeV for $p+^{15}$N decay) and the electromagnetic decay branching ratio becomes relatively very small.

We hope that new accurate measurements of the cross section of  $^{16}$O($e,e^{\prime}$)$^{16}$O(12.53 MeV, 12.97 MeV, $2^-$) and the branching ratios of $^{16}$O(12.53 MeV, 12.97 MeV, 2$^-$) decaying to the $p$, $\alpha$ and $\gamma$ channels  will be performed in the near future at the low-energy  electron accelerators ($E_e=30$-100 MeV), at MESA accelerator~\cite{MESA}  or at the ULQ2 facility at the Research Center for Electron-Photon Science (Tohoku University)~\cite{Suda}, so that the prediction of both the CC/NC neutrino-oxygen cross sections for 12.53 MeV and 12.97 MeV ($2^-$) can be accurate to a level of 10\% or less.    

\section*{Acknowledgment}
This work was partially supported by JSPS Grant-in-Aid for Scientific
Research (No. JP19K03855, JP20K03973, No. JP20K03989, and No.
JP24K07021) and also by JSPS Grant-in-Aid for Scientific Research on
Innovative Areas "Unraveling the History of the Universe and Matter
Evolution with Underground Physics" (No. JP19H05802 and No. JP19H05811), 
Grant-in-Aid for Transformative Research Areas "The creation of
multi-messenger astrophysics" (No. JP24H01817), 
and Grant-in-Aid for Transformative Research Areas "Investigation on the
Origin and Evolution of Matter in the Universe by Extremely Rare Events:
Frontier of Creating a New Insight on the Matter in the Universe" (No.
JP24H02245).

%%\nocite{*}

\end{document}